\documentstyle[aps,preprint,prb]{revtex}
\begin{document}


\title{Phase Diagram of the 1D Anderson Lattice} 
\author {M.\ Guerrero}
\address{
Department of Physics and Astronomy, 
University of California, Irvine, CA 92717} 
\author{ R.M.\ Noack} 
\address{Institut f\"ur theoretische Physik \\ Universit\"at W\"urzburg,
Am Hubland, 97074 W\"urzburg, Germany }
\date{ \today}

\maketitle
\begin{abstract}
We map out the phase diagram of the one--dimensional Anderson lattice
by studying the ground state magnetization 
as a function of band--filling using the density matrix
renormalization group technique. 
For strong coupling, we find that the quarter--filled system
has an S=0 ground state with strong 
antiferromagnetic correlations.
As additional electrons are put in,
we find first a ferromagnetic phase,
as reported by M\"{o}ller and W\"{o}lfle,
and then a phase in which the ground state has total spin
$S=0$.
Within this $S=0$ phase, 
we find RKKY oscillations in the spin--spin correlation functions.
\end{abstract}



\section{Introduction}

In recent years, heavy fermion materials have attracted a lot of interest,
from both the experimental and theoretical point of view.
These systems, usually rare earth 
or actinide compounds, show a variety of  unusual properties. 
At high 
temperatures ($T = 100 K$),  they behave as metals with weakly interacting
magnetic moments. 
When the temperature is lowered, their behavior is
consistent with the development of a narrow band of conduction electrons
with very large effective masses $m^{*}$, up to two or three orders
of magnitude larger than the bare electron mass \cite{hess}. 

The Anderson lattice Hamiltonian is believed to contain the essential physics
needed to describe the low temperature properties of heavy fermion materials. 
It considers a localized orbital at each lattice site that hybridizes with an 
extended band of conduction electrons. 
Double occupation 
of the localized orbital is penalized by a strong Coulomb repulsion $U$. 

Heavy fermions systems exhibit different kinds of ground states: 
antiferromagnetic, superconducting, paramagnetic or semiconducting
\cite{hess}.
Therefore, it is important to 
investigate the magnetism of the ground state of the Anderson lattice as a
function of the band--filling. 

Previous studies of this model have 
shown somewhat contradictory 
results regarding 
the magnetism of the ground state. 
Using the Gutzwiller approach, Rice and
Ueda \cite{Rice} studied the 
$U=\infty$ case in which doubly occupied states of the localized
orbital are forbidden. 
They found that when the
energy of the localized orbital is well below the Fermi surface, 
there is always a ferromagnetic instability (assuming no orbital 
degeneracy). 
However, they only considered uniform magnetic states in their solution. 
In contrast, the standard
mean--field slave boson treatment of the problem \cite{Riseborough} 
gives a paramagnetic 
solution for any filling in the $U=\infty$ case. 
Reynolds {\em et al}. \cite{Reynolds} reformulated the Gutzwiller approach
using the Kotliar and Ruckenstein slave boson treatment. 
They also found that a large region of the parameter space has a
ferromagnetic ground state, but they concluded 
that the Gutzwiller solution may be too biased towards the magnetic state.

M\"{o}ller and W\"{o}lfle \cite{Moller} used the Kotliar and 
Ruckenstein slave boson 
treatment to study the one--dimensional Anderson lattice. 
They concentrated
on the symmetric case \cite{explain1} in which the energy of the
localized orbital $\varepsilon_{f}$ is $-U/2$, and allowed for the
possibility of non-uniform magnetic states. 
They found that in the strong--coupling case
(large $U$) near quarter--filling there is a very narrow antiferromagnetic 
region. 
As they increased the filling they found a transition to a 
ferromagnetic state, and for even larger fillings they found a ground 
state magnetization with an incommensurate wave vector $q$.
The wave vector $q$ increases with filling and reaches $\pi$
for the half--filled system, corresponding to antiferromagnetic order.

There are also some rigorous results regarding certain special cases.
It has been shown that the ground state of the symmetric Anderson
lattice Hamiltonian is a singlet 
in the half--filled case \cite{Ueda} and has short--range antiferromagnetic 
correlations \cite{Tian}. 
Also, when the number of electrons is equal 
to the number of sites plus one (quarter--filling with one additional
electron), the ground state 
was shown to be ferromagnetic for sufficiently large $U$ 
in Ref.\ 9. 

All the methods described above rely on some approximation scheme to
solve the Hamiltonian.
For example, in
the slave boson techniques, a set of auxiliary bosons is introduced, 
in addition to the original fermions. 
In order to eliminate the non-physical 
states of the enlarged Fock space, it is necessary to impose constraints on the
boson operators.
However,
within a mean--field treatment, the constraints are not
satisfied at each lattice site but only on average for the system as a whole.
In the Gutzwiller approximation, the strong correlations
are taken into account
by renormalizing the hybridization matrix element by a factor that depends 
on the spin and on the average number of $f$--electrons per site.

In this work we use the density matrix renormalization group (DMRG) method 
\cite{White} to study the phase diagram of the one--dimensional
Anderson lattice model. 
The method gives quite accurately the properties of the exact ground
state and low--lying excited excited states on a finite cluster, but
for larger lattice sizes than, for example, Lanczos exact
diagonalization calculations.
The advantage over the analytic studies mentioned above is
that the DMRG takes into account quantum fluctuations, whereas the
analytic methods described above treat the system within mean--field 
approximations.
In the past, most numerical studies of the one--dimensional 
Anderson lattice have been limited to the symmetric half--filled case.
Here we consider fillings between quarter--filling and half--filling. 
We investigate the symmetric case using chains of 8 and 16 sites and the 
$U=\infty$ case with lattices of 8 sites. 
Our results in the strong--coupling
regime are in good agreement
with Ref.\ 5. 
Near quarter--filling we find an $S=0$ ground state.
As electrons are added we find first a ferromagnetic region, and
then once again an $S=0$ ground state for still larger fillings. 
In order to determine the nature of the magnetic correlations in the
phases with $S=0$, we examine the spin--spin correlation function.

Our results are consistent with the rigorous results described above
and also with the phase diagram obtained in a numerical study of the
Kondo lattice model \cite{Tsunetsugu}. 
Since the symmetric Anderson Hamiltonian can be mapped into
the Kondo Hamiltonian \cite{Schrieffer} when the hybridization between
the $f$--band and the conduction band is small compared to $U$, 
the phase diagrams should be similar in this regime.

This work is organized as follows. 
We briefly describe the one--dimensional
Anderson lattice Hamiltonian and discuss some of its properties in
section \ref{PERANDSEC}.
In section \ref{RESULTS} we present the numerical results. 
We study chains of 8 and 16 sites for the case in which
$\varepsilon_{f}=-U/2$ in section \ref{SYMCASE}, and 
draw a phase diagram based on the total spin of ground state and the
nature of the spin--spin correlation functions. 
In section \ref{UINFCASE} we construct the
phase diagram for $U=\infty$ using results on chains of 8 sites. 
Our conclusions are given in section \ref{CONCLUSION}.

\section{\label{PERANDSEC} The Periodic Anderson Hamiltonian}

We consider the standard periodic Anderson Hamiltonian in one dimension:
\begin{equation}
H=-t\sum_{i \sigma} (c^{\dagger}_{i \sigma}   c_{i+1 \sigma} +
                     c^{\dagger}_{i+1 \sigma} c_{i\sigma} )
  + \varepsilon_{f} \sum_{i \sigma} n^{f}_{i \sigma} 
  + U \sum_{i} n^{f}_{i \uparrow}n^{f}_{i \downarrow}
  + V \sum_{i \sigma} (c^{\dagger}_{i \sigma} f_{i \sigma} + 
                      f^{\dagger}_{i \sigma}  c_{i \sigma} )     
\label{eq:Hamiltonian}
\end{equation}
where $c^{\dagger}_{i \sigma}$ and $c_{i\sigma}$ create and annihilate
conduction 
electrons with spin $\sigma$ at lattice site $i$, and  $f^{\dagger}_{i
\sigma}$ and $f_{i \sigma}$ create and annihilate local $f$--electrons. 
Here $t$ is the hopping matrix element for conduction electrons between
neighboring sites, 
$\varepsilon_{f}$ is the energy of the localized $f$--orbital,  
$U$ is the on--site Coulomb
repulsion of the $f$--electrons, and $V$ is the on--site hybridization
matrix element 
between electrons in the $f$--orbitals and the conduction  band.
For simplicity, we neglect orbital degeneracy. 
We denote the number
of electrons by $N_{el}$, and $N$ is the number of sites in the lattice. 
Since there are two electronic orbitals in each site, the quarter--filled case 
corresponds to $N_{el}=N$ and the half--filled case has $N_{el}=2N$.

For $U=0$ this Hamiltonian can be exactly diagonalized in momentum
space, yielding two hybridized bands with energies $\lambda_{k}^{\pm}$:
\begin{equation}
\lambda_{k}^{\pm}=\frac{1}{2}\left[(\varepsilon_{f}-2t\cos(ka)) \pm
\sqrt{(\varepsilon_{f}+2t\cos(ka))^{2}+4V^{2}}\right],
\end{equation}
where $a$ is the lattice constant.
Therefore, when the number of electrons is between quarter--filling 
and half--filling, the lower
band is occupied but the upper band is always empty, and
the ground state is paramagnetic for any filling.

Now consider the case when the $f$--level is well below the conduction
band and the Coulomb repulsion $U$ is large. 
With no hybridization ($V=0$), the ground state at quarter--filling
has one electron at each $f$--site and there is degeneracy in the spin
configurations. 
When  $V>0$, exchange interactions remove this degeneracy. 
It can be shown using perturbation theory 
that the effective interaction between neighboring sites
favors antiferromagnetic ordering of neighboring $f$--electrons 
\cite{Moller,Bastide}. 
The relevant exchange process is sixth order
and involves an $f$--electron hopping to the conduction band, 
then to a nearest neighbor conduction site and then into the
$f$--orbital on that site. 
In the intermediate state, the $f$--orbital is doubly occupied, which is
only possible if the spins of the electrons are opposite.
This leads to an effective antiferromagnetic interaction.

When the filling is increased slightly, the additional electrons go into the
conduction orbitals because of the strong Coulomb repulsion $U$ in
the $f$--orbitals. 
In this case, there is an on--site antiferromagnetic
correlation between the electron in the conduction orbital and the one in 
the $f$--orbital, favoring a local singlet. 
To optimize the kinetic energy of the conduction electrons, it is
favorable for the $f$--electrons to have their
spins oriented in the same direction \cite{Moller,Yanagisawa}. 
Therefore, if there are $N_{c}$ conduction electrons compensating the
$f$--spins, one expects a ferromagnetic ground state with $S=(N-N_{c})/2$. 
When this value of $S$ is realized, we will call it {\em complete}
ferromagnetism. 
If the value of $S$ we find is smaller
than the complete value, but still greater than the minimum (0 or
1/2), then we will refer to it as {\em incomplete} ferromagnetism,
meaning that not all the uncompensated $f$--electrons are aligned.
These two effects give rise to a competition between  
ferromagnetic and antiferromagnetic ordering near quarter--filling
\cite{Moller}.

On the other hand, 
when the filling is further increased in the strong--coupling case,
the interaction between $f$--electrons, mediated 
by the Fermi sea, starts to play an important role. 
This is the well--known
RKKY \cite{Ruderman} interaction that induces correlations with 
wavevector $q=2k_{F}$ between the $f$--electrons, where $k_{F}$ is the
Fermi wavevector of the non-interacting ($V=0$) Fermi sea of
conduction electrons.
 
For simplicity, we concentrate here on two particular cases of the
Anderson Hamiltonian: the symmetric case\cite{explain1}  
in which $\varepsilon_{f}=-U/2$ and the $U=\infty$ case. 
This reduces the number of independent Hamiltonian parameters by one. 
In the symmetric case, strong coupling (large $U$) means that the
$f$--level is far below the conduction band. 
Therefore we expect to find a competition between antiferromagnetic and 
ferromagnetic correlations near quarter--filling and to find RKKY 
correlations for larger fillings.
For small $U$, we expect a paramagnetic ground state.
In the $U=\infty$ case we set the $f$--level $\varepsilon_{f}$ to
be less than or equal to 0.
Again, when $\varepsilon_{f}$ falls below the conduction band, we expect 
competition between antiferromagnetic and ferromagnetic correlations near 
quarter--filling, and RKKY interactions near half--filling.

\section{\label{RESULTS} Results}

\subsection{\label{SYMCASE} The Symmetric Case }

We first consider the symmetric case, $\varepsilon_{f}=-U/2$. 
We fix $t=0.5$, $V=0.375$ and vary $U$ from 0 to 6 
(all energies are in units of $2t$, which is half the bandwidth). 
This choice of
parameters allows us to do a quantitative comparison with Ref.\ 5. 
We use the DMRG technique\cite{White}
to find the energies and equal--time correlation functions of the
ground state and low--lying states on finite lattices.
While this technique gives energies that are, in principle,
variational, it has proven to give quite accurate results for 1D
quantum lattice systems.
The method provides a controlled way of numerically diagonalizing a
finite system within a truncated Hilbert space.
One can increase the accuracy by increasing the number of states kept,
and can examine the convergence with the number of states.
Here we typically keep up to from 150 to 200 states per block, although 
in the numerically more difficult cases, such as the calculation of
the correlation functions for the 16 site chains, we keep up to 400 states.
Truncation errors, given by the sum of the density matrix
eigenvalues of the discarded states, vary from $10^{-5}$ in the worse
cases to $10^{-9}$ in the best cases.
This discarded density matrix weight is directly correlated with the
absolute error in the energy.
Since the method is most accurate for a given amount of computational
effort when the system has open boundary conditions (i.e. no
nearest--neighbor connection between site 1 and site $N$), 
we apply open boundary conditions here.

Within the DMRG method, we fix the number of electrons $N_{el}$ and
the $z$--component of the total  
spin of the system $S_{z}$ and find the ground state within this subspace.
In order to determine the nature of the ground state, we would like to
determine the total spin, $S$.
For a ground state of a given $S_{z}$, there are several possible values of
the total spin $S$ ($S_{z} \le S \le N_{el}/2$). 
In order to establish the value of $S$, 
we calculate the mean value of the operator ${\bf S}^{2}$ in the ground
state with the lowest possible $S_{z}$ (0 or 1/2 according to whether
$N_{el}$ is even or odd). 
In this way we can be sure that we are considering
all the possible values of $S$.
Since $\langle {\bf S}^{2}\rangle=S \dot (S+1)$ (setting  $\hbar=1$), 
we can deduce the value of $S$.
For example, for 8 sites with $U=4$ and $N_{el}=9$, we
obtain $\langle {\bf S}^{2} \rangle =15.748$ for the $S_{z}=1/2$
ground state, implying $S=7/2$.

In some cases, states with different values of $S$ can be close in energy.
When this happens, the wave function obtained
for the ground state with a given $S_z$ can be composed of a mixture
of states with 2 or
more $S$ values, rather than having a definite value of $S$. 
This occurs mainly for longer chains ($N \ge 16$), for which the
numerical accuracy is lower and the states are closer together in energy.
In these cases, although we cannot immediately 
determine the value of $S$, we can conclude that it is not the smallest
possible value. 
We can then study states with higher 
values of $S_{z}$, for which the Hilbert space
is smaller (there are fewer values of $S$ allowed) and therefore there is
less mixing. 
Also, since we keep the same number of states in a smaller
Hilbert space, the numerical accuracy is higher.
For example, on a 16 site lattice with $U=2$ and $N_{el}=22$, we obtain 
$\langle {\bf S}^{2}\rangle =25.38$ for the lowest $S_{z}=0$ state.
This indicates the $S$ is 
likely to be higher than  $3$ but it could be either $4$ or $5$. 
We then consider the lowest energy $S_{z}=2$, $3$, $4$ and $5$
states, and obtain
$\langle {\bf S}^{2} \rangle =26.39$, $28.35$, $29.97$ 
and $30.00$ respectively. 
The energies in all cases are degenerate to within the estimated
accuracy of the calculation.  
Therefore, we conclude that $S=5$ for this case.

In Fig.\ 1 we present our results for the spin $S$ of the ground state of
the 8 site chain, showing the number of electrons on the horizontal
axis and $U$ on the vertical axis.
At quarter--filling ($N_{el}=N=8$), we find the ground
state always has $S=0$. 
Also, for $U=0$ or $U$ small, we find that the ground state is 
paramagnetic at all fillings, as predicted by the qualitative picture
given in section \ref{PERANDSEC}.
For $U \ge 2$, we find a narrow ferromagnetic region slightly above
quarter--filling (enclosed with a solid line as a guide to the eye). 
We circle the cases of {\em complete} ferromagnetism as defined in the
previous section. 

For larger fillings,  we find an $S=0$ ground state for all couplings $U$.
However, when the number of electrons is odd we obtain $S=3/2$ and not
$S=1/2$ as one would expect.
We attribute this to a finite size effect for the following reason:
if we consider chains with 16 sites with the same density of
electrons, (for example, $U=4$ with 22,\, 26,\, 30 electrons),
we find $S=0$ in the ground state. 
This alternation of $S=0$ and $S=3/2$ states was also observed in 
Ref.\ 13 
in the context of the phase diagram of Kondo
lattice model.
The $S=3/2$ state appears when there is an 
odd number of electrons in the conduction band, so that one of the conduction 
energy levels has a single electron. 
The $f$--electrons will then interact 
mainly with the single unpaired electron and will tend to align
ferromagnetically \cite{Tsunetsugu}. 
Roughly speaking, for an $f$--electron to interact with one electron of 
the doubly occupied conduction band and produce a spin flip with energy 
gain $J_{eff}$, one conduction electron needs to hop to a higher energy 
level. 
When the effective Kondo coupling,
$J_{\rm eff}$ (given by the Schrieffer-Wolff transformation 
\cite{Schrieffer}), is less than the spacing of the conduction electron 
energy levels, the $f$--electrons can only couple with the unpaired
conduction electron.
In fact, as $J_{\rm eff}$
decreases, this effect becomes more important and, presumably for
$J_{\rm eff}$ 
small enough, the ground state should have
the maximum value, $S=(N-1)/2$. 
However, in the infinite
system there is no finite separation between conduction energy levels, and
the ground state should be paramagnetic for any value of $J_{\rm eff}$.

In order to better understand the nature of the correlations in the
antiferromagnetic phase at quarter--filling and the transition to the
ferromagnetic phase as the filling is increased, we have also carried
out calculations on a 16 site lattice.
At quarter--filling, the ground state is $S=0$ for all the $U$ values
we considered, but as $U$ increases there is an onset of short--range
antiferromagnetic correlations.
In Fig.\ 2 we plot the $f$--spin--$f$--spin correlation function at
quarter--filling for different values of $U$.
For $U=0$, the correlations are very small and always negative. 
As $U$ increases, they alternate in sign and increase in amplitude.
This result is consistent
with Ref.\ 5 
which found a narrow antiferromagnetic
region near quarter--filling. 
For the fully interacting system in 1D, treated exactly by the DMRG,
quantum fluctuations destroy the long--range antiferromagnetic
correlations found in the mean--field slave boson calculations, but
short--range antiferromagnetic correlations remain.

For $U=2,3,4$ and $6$, we map out the extent of the ferromagnetic
phase by increasing $N_{el}$ until the ground state becomes
paramagnetic.
We plot the resulting phase diagram in Fig.\ 3.
Here ``C'' denotes the states with complete
ferromagnetism ($S=[N_{el}-N_{c}]/2$) and ``I'' denotes the states
with incomplete ferromagnetism ($S < [N_{el}-N_{c}]/2$ but larger
than the lowest possible value).
The states of incomplete ferromagnetism in the the boundary region
between the ferromagnetic and antiferromagnetic phases suggest
that the ferromagnetic order parameter may go to zero continuously,
implying a second order phase transition.

For the $U=6$ and $U=4$ points with $N_{el}=20 $,
the $U=3$, $N_{el}=22$, and the $U=2, N_{el}=18$ points in Fig.\ 3, 
the difference in energy between the $S=0$ and $S=1$ states is of the
order of the numerical accuracy, making it hard to determine the total
spin of the ground state. 
However, we include these points in the paramagnetic region
because an S=1 ground state, although still ferromagnetic, indicates a
very strong suppression of the ferromagnetism,
and because the ground state is paramagnetic at the same
parameters and average fillings in the 8 site chain. 
Also, for $U=2$, $N_{el}=19$ and $N_{el}=23$ the states are also very close 
in energy and it is very hard to establish the value of $S$ in the ground 
state. However, we can establish that $S$ is greater than $1/2$
and that is smaller than $(N-N_{c})/2$ so we list these points as incomplete
ferromagnetism.

By comparing the results of 8 and 16 site chains for the same density
of electrons $n=N_{el}/N$, one can see that the cases 
of complete ferromagnetism are always consistent. 
The incomplete ferromagnetism  is systematic in the sense that for a given
electron density, the incomplete ferromagnetism appears in both 8 and
16 site chains. 
However, the value of $S$ does not necessarily scale 
with the number of sites.
For example, for $U=3$ and $n=1.125$ we find $S=3/2$ for 8 sites and
$S=5$ for 16 sites.

For $U=4$ and $N=16$, we examine the spin--spin correlation functions at
larger fillings ($N_{el}=24$, 28, 32).
We calculate $C(q)$, the Fourier transform of
$<S_{z}^{f}(R)S_{z}^{f}(0)>$ where R is the distance in units of the 
lattice constant \cite{explain}, 
for $N_{el}=24, \, 28$ and $32$. 
The continuous Fourier transform is calculated by zero padding the
function $<S_{z}^{f}(R)S_{z}^{f}(0)>$ for $R > N$. 
In order to reduce spurious
high frequency oscillations introduced by cutting off the real--space
correlation function at the open boundaries,
we window the data using a Bartlett windowing function
\cite{numrecipes} over the interval $0 < R < N$ before transforming.

We plot $C(q)$ in Fig.\ 4 and
we see that for each case there is a peak in $C(q)$
at $q=2k_{F}$ where $k_{F}$ is the Fermi wave vector of the non-interacting 
($V=0$) conduction band ($k_{F}=\pi/4$, $3\pi/8$ and $\pi/2$ for $N_{el}=24$,
$28$ and $32$, respectively). 
This form is characteristic of RKKY oscillations which are important
in this $S=0$ regime.(The peaks for $N_{el}=24$ and $28$ are 
slightly shifted from the exact value of $2k_{F}$ , the shift is 
roughly 2 percent)

We can compare our results with those of Ref.\ 5 
in which
the symmetric one--dimensional Anderson lattice was studied for the
strong--coupling case 
using the Kotliar and Ruckenstein slave boson technique (the results for
$U \ge 2.5$ are in their Fig.\ 9).  
In their antiferromagnetic region we find an $S=0$ ground state with 
short--range antiferromagnetic correlations that increase in magnitude and 
range as $U$ increases. 
The parameter regimes in which we find complete ferromagnetism and incomplete
ferromagnetism fall within the limits of 
their ferromagnetic region with the exception of our point at
$U=6, \, N_{el}=19$ which lies in their paramagnetic region.
We find incomplete ferromagnetism in the ground state at this point. 
This discrepancy could be due to the finite size effect described
earlier in which there is a tendency  
towards a ferromagnetic state in the cases
with an odd number of electrons in the conduction band.
The ferromagnetism is always complete in Ref.\ 5 
presumably due to the mean--field nature of their calculation.
In contrast, we find a region of incomplete ferromagnetism in the boundary
between the antiferromagnetic and ferromagnetic regions that suggests
that the phase transition may be second order.
At half--filling they find an antiferromagnetic ground state 
(in the strong--coupling regime). 
As they decrease the filling, the magnetic
wave vector decreases linearly with the doping concentration from its
value $q=\pi$ at half--filling.
We associate this
with the RKKY correlations with wavevector $2k_{F}$ that we find in a wide 
region below half--filling, since $k_{F}$ is proportional to the electron 
density in one dimension.
We therefore find that our phase diagram is in good overall agreement
with that of M\"{o}ller and  W\"{o}lfle \cite{Moller}.

\subsection{\label{UINFCASE} The Asymmetric $U=\infty$ Case}

We also study the asymmetric Anderson model at $U=\infty$, again fixing 
$V=0.375$. 
We vary the position of the $f$--level $\varepsilon_{f}$ from 0 to
$-2.5$ and study 8 site chains, keeping 100 states per block for the
small $|\varepsilon_{f}|$ cases and up to 250 states per block for the larger
$|\varepsilon_{f}|$.
In Fig. 5, we tabulate the total spin $S$ of the ground state as a
function of the number of electrons $N_{el}$ (horizontal axis) and
the absolute value of $\varepsilon_{f}$ (vertical axis).
We consider $\varepsilon_{f} \le 0$ only. 

There is a clear resemblance between Fig.\ 5 and Fig.\ 1.
As before, at exactly quarter--filling the ground state has $S=0$ and
there are increasing antiferromagnetic correlations as the $f$--level
falls below the bottom of the conduction band. 
There is a narrow ferromagnetic region near quarter--filling
and then a paramagnetic region at larger fillings.  
The ferromagnetic region starts roughly where the $f$--level falls below the
conduction band (Kondo regime). 
For small values of $\varepsilon_{f}$
(mixed--valence regime), we find a paramagnetic state at all fillings.
This is in contradiction with
the Gutzwiller result that predicts that there will always be a 
ferromagnetic instability at any filling.
At quarter--filling, antiferromagnetic correlations prevail,
and at larger fillings, there is a region in which the ground state
has $S=0$.  
In this region, RKKY interactions presumably dominate the
magnetic correlations, as in the symmetric case.
In a previous study, it was shown that for the half--filled system,
RKKY correlations are important in the Kondo regime but are strongly
suppressed in the mixed--valence regime \cite{us}.

In the mixed--valence
region there is no ferromagnetism at any filling, in agreement
with the slave boson mean field approach. 
However, the slave boson treatment 
predicts a paramagnetic state for any value of $\varepsilon_{f}$. 
This suggests that the slave boson description is appropriate for the
mixed--valence case, but breaks down in the Kondo regime.

\section{\label{CONCLUSION} Conclusions}

We constructed the phase diagram of the one--dimensional Anderson lattice 
using the density matrix renormalization group technique. 
The results are summarized in Fig. 3.
We considered the symmetric case with $\varepsilon_{f}=-U/2$ and the
asymmetric case with $U=\infty$. 
In the symmetric case for large $U$ we
found an $S=0$ ground state with short--range 
antiferromagnetic correlations at quarter--filling, that increase as 
$U$ increases.
At slightly larger fillings,
there is a transition to a ferromagnetic state. 
The presence
of a small region of incomplete ferromagnetism in the boundary suggests 
a second order transition. 
For small values of $U$ in the symmetric
case we find, as expected, a paramagnetic state at all fillings. 
For small values of $|\varepsilon_{f}|$, the phase diagram of the
$U=\infty$, asymmetric case is quite similar.

In the strong--coupling limit in the symmetric case, we compared our
results with Ref.\ 5 
which studied the one--dimensional
Anderson lattice using the Kotliar--Ruckenstein slave boson
approach.
We found good qualitative agreement with their results. The ferromagnetic
region is the same in both cases. 
However, we find incomplete ferromagnetism
in the boundary with the $S=0$ region near quarter--filling, in
contrast to the sharp transition found in M\"{o}ller and W\"{o}lfle's
work. 
Also, where they find long--range 
antiferromagnetic order, we obtain short--range antiferromagnetic
correlations. 
This can be attributed to the presence of quantum fluctuations that are 
not taken into account in their treatment.
In the strong--coupling case, our phase diagram is 
consistent with the phase diagram of the Kondo lattice Hamiltonian
\cite{Tsunetsugu} in the small $J$ region. 

In the $U=\infty$ case, our results agree with the predictions of the 
standard slave boson mean field approach \cite{Riseborough} only for 
small values of $|\varepsilon_{f}|$ (the mixed--valence case). 
For larger values of $|\varepsilon_{f}|$, 
the standard slave boson technique fails to predict ferromagnetism and RKKY 
correlations.

\section*{Acknowledgments}

We would like to thank Steve White and  Kazuo Ueda for helpful discussions.
We also thank Clare Yu for critically reading the manuscript.
This work was supported in part by ONR Grant No N000014-91-J-1502 and an 
allocation of computer time from the University of California, Irvine.
Some of the numerical calculations were performed at the San Diego
Supercomputer Center.

\begin{figure}

\caption{
Values of the spin $S$ for different values of $U$ and $N_{el}$
in the ground state for 8 site chains. Parameters 
are $\varepsilon_{f}=-U/2$, $t=0.5$, $V=0.375 $.
There is a narrow ferromagnetic region near quarter--filling (enclosed by a 
solid line). Complete ferromagnetic states are circled.
}
\label{fig1}
\vspace*{0.9cm}

\caption{
The $f$--spin--$f$--spin correlation functions
versus distance $R$ apart at quarter--filling for 
$\varepsilon_{f}=-U/2$, $t=0.5$, $V=0.375 $ and different
values of $U$. 
Antiferromagnetic correlations develop as $U$ increases.
}
\label{fig2}
\vspace*{0.9cm}

\caption{
The phase diagram for the 1D Anderson lattice combining
results of chains of 8 and 16 sites. Parameters 
are $\varepsilon_{f}=-U/2$, $t=0.5$, $V=0.375 $, $n=N_{el}/N$.
Here ``C'' denotes {\em complete} ferromagnetism and ``I'' {\em incomplete} 
ferromagnetism as defined in the text. 
}
\label{fig3}
\vspace*{0.9cm}

\caption{
The Fourier transform of the
$f$--spin--$f$--spin correlation 
functions for $\varepsilon_{f}=-U/2$, $t=0.5$, $V=0.375$, $U=4$ and
different fillings. The peaks appear at $q=2k_{F}$.
}
\label{fig4}
\vspace*{0.9cm}

\caption{
Values of the spin $S$ for different values of 
$|\varepsilon_{f}|$ and $N_{el}$
in the ground state for $U=\infty$, $t=0.5$, $V=0.375$ and chains with
8 sites. 
There is a narrow ferromagnetic region near quarter--filling. 
}
\label{fig5}

\end{figure}


\begin{thebibliography}{99}
\bibitem{explain1} Note that technically speaking the symmetric case 
corresponds to $\varepsilon_{f}-\mu = -U/2$, where $\mu$ is the chemical 
potential 
of the non-interacting conduction electrons which varies as a function of the 
filling. The case $\varepsilon_{f}= -U/2$ corresponds to the symmetric case at 
half--filling for which $\mu=0$.
\bibitem{explain} The correlation functions 
$ \langle S_{z}^{f}(R)S_{z}^{f}(0) \rangle$ for
a certain distance $R$ were obtained by averaging over all
pairs of sites separated by a distance $R$. 
This procedure averages out the effects of local density fluctuations
due to the open boundaries and results in a correlation function 
that is close to the infinite system one until $R$ gets to within a few
lattice spacings of the system size.
\end{thebibliography}
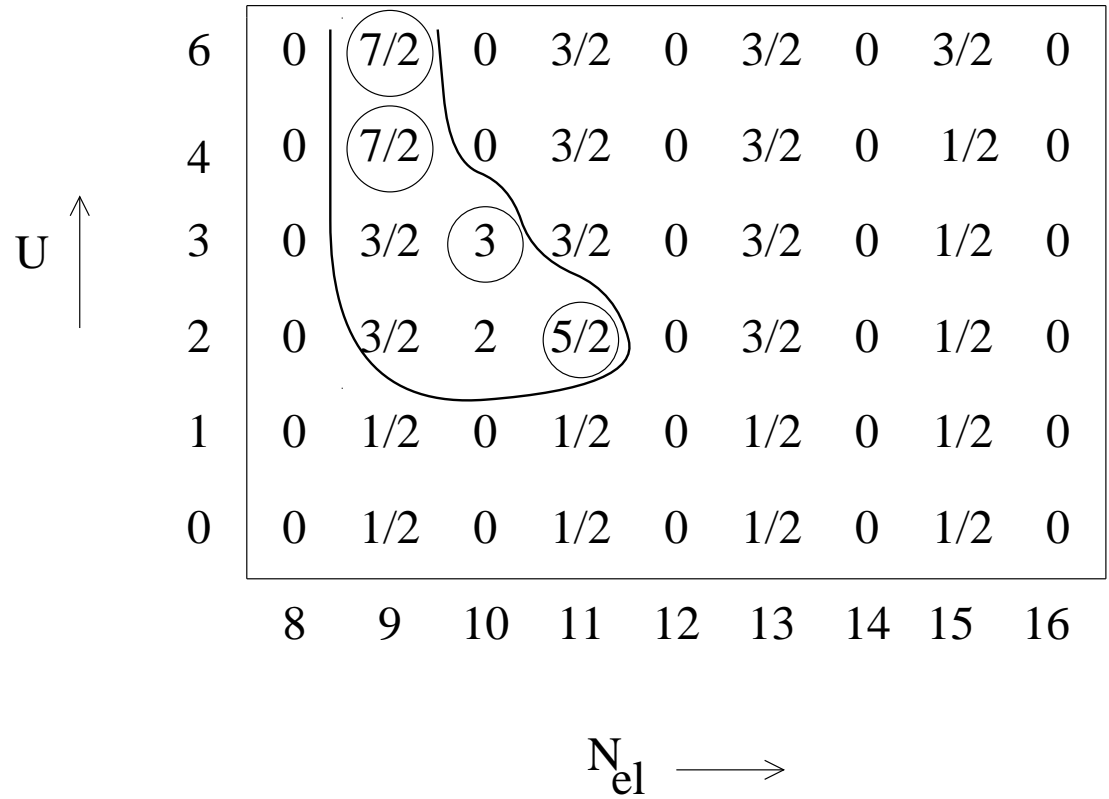
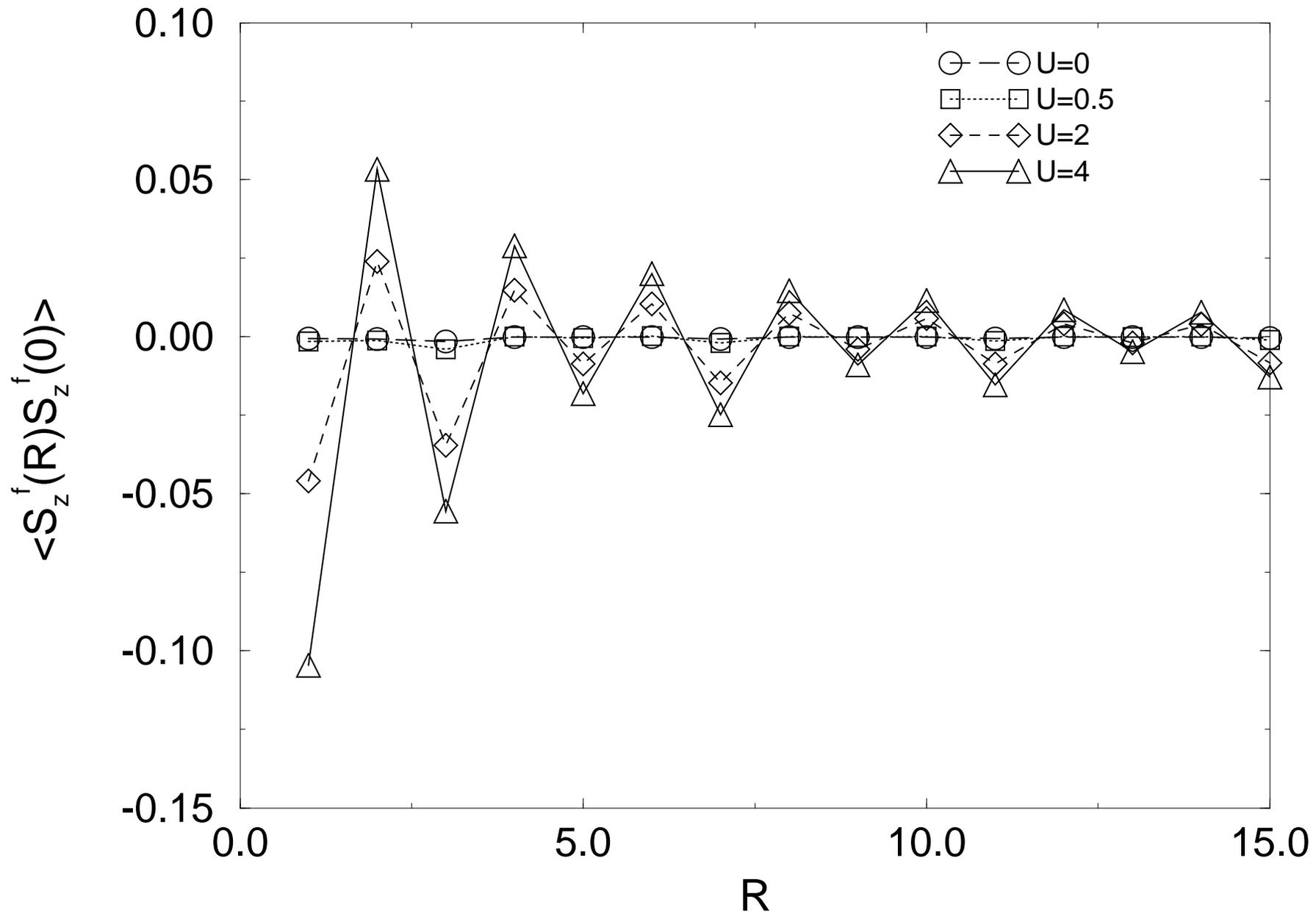
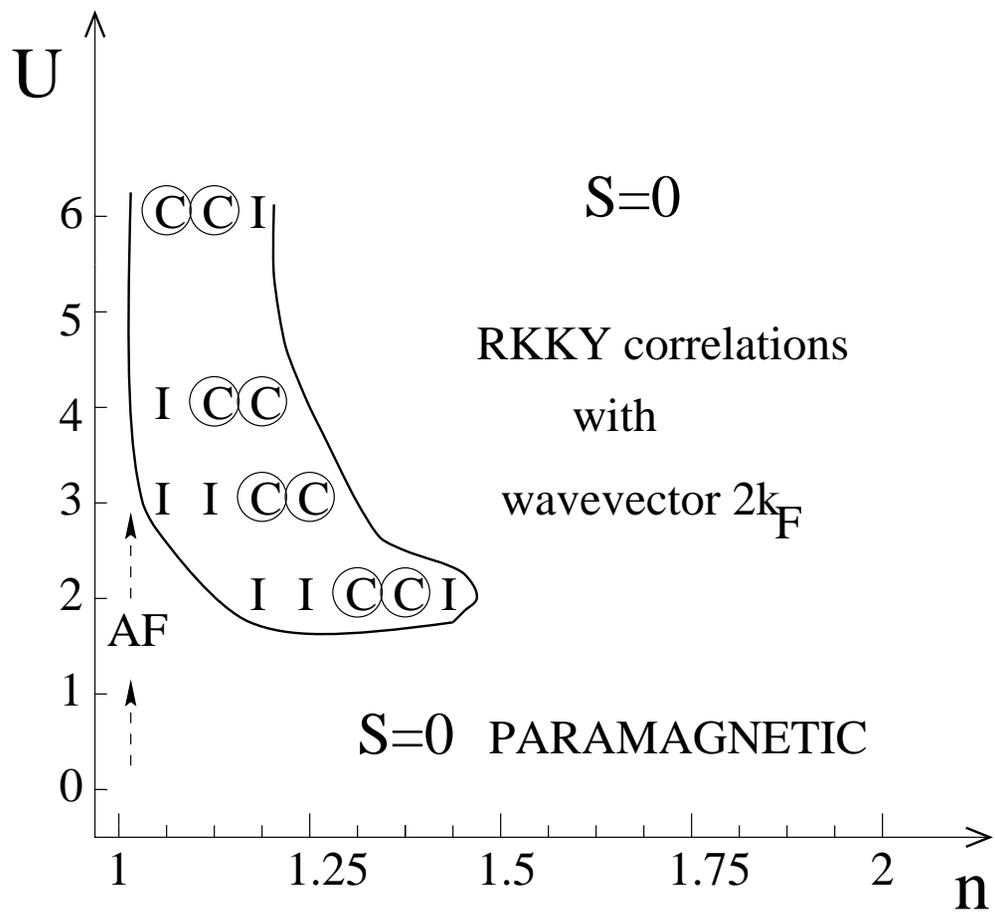
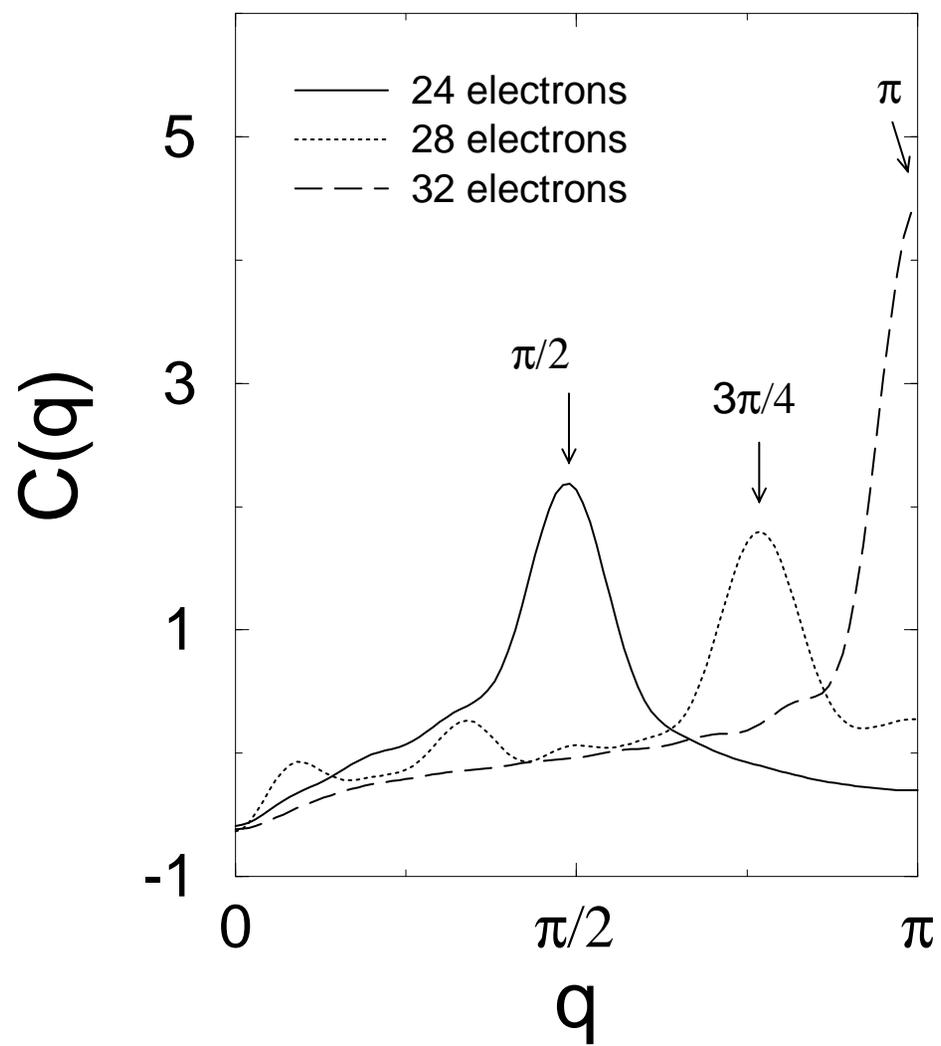
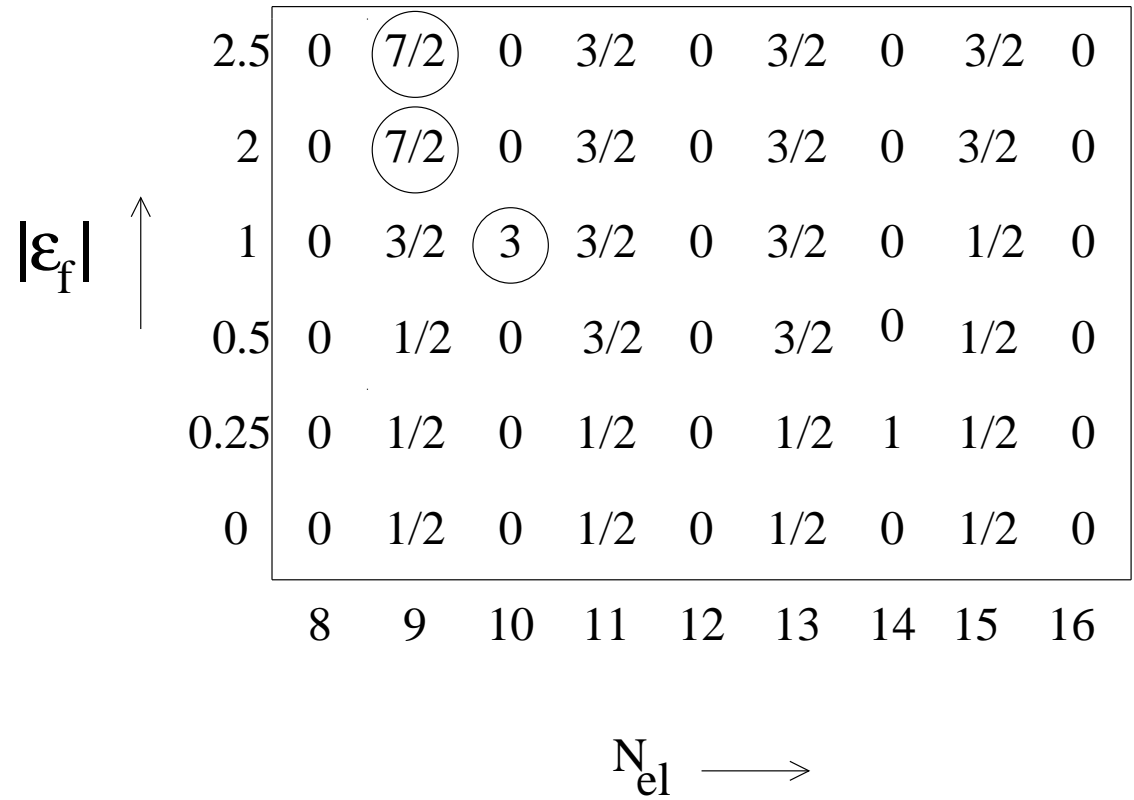
\end{document}